\documentstyle[12pt]{article}
\begin{document}
\baselineskip 21pt  plus .1pt minus .1pt
\pagestyle{empty}
\begin{center}
{\large\bf{Recent Neutrino Experiments and Their Consistency In
An Extended Harvard Model}}
\end{center}
\vskip .50in
\begin{center}
{\bf Ambar Ghosal and Asim K. Ray}
\vskip .05in
{Department of Physics\\
Visva-Bharati University\\
Santiniketan 731 235, India}
\end{center}
\vskip .5in
\begin{center}
{\bf Abstract}
\end{center}
\vskip .25in
We demonstrate that the solar and atmospheric neutrino data
as well as the recent result of the LSND experiment cannot be satisfied
simultaneously with three light neutrinos if we consider the
mass degeneracy for two neutrinos in the context of an extended
Harvard Model based on the gauge group $SU(2)_{qL}\times
 {SU(2)}_{lL}\times {U(1)}_Y$ with $S_3\times Z_4$ discrete symmetry.
Assuming two different
representation contents under $S_3\times Z_4$ symmetry for
pairwise neutrinos and the lone  neutrino ($\nu_e$ and $\nu_\mu$
transforming as a doublet and $\nu_\tau$ as singlet or $\nu_e$ and
$\nu_\tau$ as a doublet and $\nu_\mu$ as singlet) the present model admits
neutrino
masses of the order of 2.8 eV and can fit either solar and
atmospheric neutrino data or the LSND and solar neutrino data.
\newpage
\pagestyle{plain}
\setcounter{page}{2}
\noindent
Neutrino mass is one of the key issues of the present day
particle physics. Although there is no principle which dictates
that the neutrino mass to be zero, the Standard Model of
particle physics assumes zero mass for three generations of
neutrinos. Recent experiments on the solar neutrino deficit [1],
atmospheric neutrino anomaly [2], the excess of $\bar{\nu_\mu}-
\bar{\nu_e}$ events observed recently by the Liquid Scintillator
Neutrino Detector (LSND) experiment [3] and the need for a
cosmological hot dark matter component [4] suggest that the
neutrinos have non-zero mass of the order of a few eV.
Wolfenstein [5] has pointed out that the LSND result combined
with the Zee model [6] leads to the interesting predictions that
there are two neutrinos almost degenerate with masses of
interest for cosmology and a large neutrino oscillation signal
should be seen on either the atmospheric neutrinos or the solar
neutrinos. In other words, the atmospheric neutrino anomaly and solar neutrino
deficit can be explained due to $\nu_\mu\rightarrow \nu_\tau$
and $\nu_e\rightarrow \nu_\tau$ oscillations respectively. Ma
and Roy have proposed a model [7] of four light neutrinos
$\nu_e, \, \nu_\mu,\,\nu_\tau$ and a singlet $\nu_s$ in the
framework of $SU(2)_L\times U(1)_Y\times Z_5$ model to explain
the recent data of the above mentioned experiments. They
conclude that neutrino oscillation can explain the solar
neutrino deficit $(\nu_e\rightarrow \nu_s)$, the atmospheric
neutrino anomaly $(\nu_\mu\rightarrow\nu_\tau)$ and the LSND
observed experiment ($\bar{\nu_\mu}-\bar{\nu_e}$). In this paper we examine
the consistency of the results of the above mentioned neutrino experiments
in the
context of an extended Harvard model based on
the gauge group $SU(2)_{qL}\times SU(2)_{lL}\times U(1)_Y$ [8]
with appropriate Higgs fields and discrete symmetry, which has been recently
 studied
to
achieve spontaneous CP violation [9]
and obtain
neutrino mass and
magnetic moment [10].\\

The model is based on the gauge group
$SU(2)_{qL}\times SU(2)_{lL}\times U(1)_Y$ with $S_3\times Z_4$
discrete symmetry.
We concentrate
on the  lepton
and  Higgs  fields  of  the
model. The ordinary leptonic fields
($l_{il},\,\nu_{iR},\,e_{iR}$,\,i=1,2,3),
spectator
fields ($E_{iL}, E_{iR}$, i=1,2,3) and the Higgs fields
$\phi_\alpha(\alpha=1,..3)$
, $\eta$ , $\Sigma$
have  the following representation contents :
$$l_{iL}(1,2,-1,1),e_{iR}(1,1,-2,1),\nu_{iR}(1,1,0,1),
 E_{iL}(1,2,-1,1),
E_{iR}(2,1,-1,1),$$ $$\phi_{\alpha}(1,2,1,0),
\eta(1,1,0,-2),
\Sigma(2,2,0,0)\eqno(1)$$
\noindent
where the digits in the parenthesis represent $SU(2)_{qL}$,
$SU(2)_{lL}, U(1)_Y$ and  Lepton number L(=$L_e + L_\mu + L_\tau$)
respectively.

The Higgs content
of the model gives rise to two step breaking of the ununified
gauge group down to $U(1)_{em}$. The
bi-doublet Higgs field $\Sigma$ breaks the ununified gauge group down to
the Standard Model and has no direct contribution to the neutrino mass matrix.
The mass matrix is generated in the model through the see-saw mechanism [11]
and in the
$(\nu_L,\, \nu_R^c)$ basis is given by \\

$$M_\nu = \pmatrix{0&m_D\cr
                   m_D^T&m_R}\eqno(2)$$
\noindent

The $\phi_\alpha$ fields break the Standard
model gauge group to the $U(1)_{em}$ and contribute to the $3\times 3$
Dirac mass matrix
$m_D$. The singlet Higgs field
$\eta$ leads to spontaneous lepton number violation (SLV) due to
its non-zero VEV and contributes to the right-handed $3\times 3$
 Majorana mass matrix $m_R$.
This is in contrary to the Zee model [6] in which explicit
violation of lepton number occurs. However, all the SLV processes
(such as $\mu\rightarrow
e\gamma, K_L\rightarrow \mu e$ etc.) are highly suppressed due to
the small mass squared differences $\Delta_{ij} = m_{\nu_i}^2 - m_{\nu_j}^2$
of neutrinos. Apart from the
electroweak symmetry breaking scale, the present model contains two
other intermediate mass scales, the ununification
symmetry breaking scale and the lepton number symmetry breaking scale.
It  is  to  be  noted  that,
the
spectator fermions are necessary for
anomaly cancellation [8].\\

We consider two different types ( Type A and Type B ) of
transformations of the
ordinary leptons under $S_3\times Z_4$ discrete symmetry.
The ordinary leptons, spectator fermions and the Higgs fields
transform under $S_3 \times Z_4$ discrete symmetry
as follows:\\
\noindent
\underline{Type A}\\
\noindent
i) $S_3$ symmetry :\\
$$(l_{1L}, l_{3L})\rightarrow 2, l_{2L}\rightarrow 1,
(E_{1L}, E_{2L})\rightarrow 2, (E_{1R},E_{2R})\rightarrow 2$$
$$(\nu_{\tau R}, \nu_{eR})\rightarrow 2, \nu_{\mu R}\rightarrow 1,
E_{3L}\rightarrow 1, E_{3R}\rightarrow 1,
(\mu_R, e_R )\rightarrow 2, \tau_R\rightarrow 1,$$
$$\phi_1 \rightarrow 1,(\phi_2, \phi_3)\rightarrow 2,
\eta \rightarrow 1,
\Sigma \rightarrow 1
\eqno(3a)$$
\noindent
ii) $Z_4$ symmetry:\\
$$(\mu_ R, e_R)\rightarrow -i(\mu_R, e_R),
\tau_R\rightarrow -i\tau_ R,
(\phi_2,\phi_3)\rightarrow i(\phi_2,\phi_3)\eqno(3b)$$
all other fields are invariant under $Z_4$ symmetry transformation.\\

\noindent
\underline{Type B}\\
\noindent
i) $S_3$ symmetry :\\
$$(l_{1L}, l_{2L})\rightarrow 2, l_{3L}\rightarrow 1,
(\nu_{\mu R}, \nu_{eR})\rightarrow 2, \nu_{\tau R}\rightarrow 1.\eqno(4)$$
The rest of the fields transform as in Type A.\\
\noindent
ii) $Z_4$ symmetry:\\
Under $Z_4$ symmetry all the fields transform similar to Type A.\\
\noindent
The   purpose   of
incorporation of $S_3$ permutation symmetry is to generate the equality
between the Yukawa couplings and VEV's of the neutrinos.
The $\phi_2$ and $\phi_3$ Higgs fields are necessary to achieve
non-degenerate charged lepton mass matrix.
The discrete
$S_3$ symmetry also protects the diagonal form of
$m_R$ as well as gives rise to some vanishing terms
in $m_D$, which simplifies the diagonalization of the entire
mass matrix and also leads to the
almost degenerate neutrino mass.
The discrete $Z_4$ symmetry prohibits $\phi_2$ and $\phi_3$
Higgs fields to couple with the neutrinos and $\phi_1$ Higgs
field to the charged leptons. Thus, the charged lepton mass
matrix becomes completely different from the neutrino mass
matrix. We now concentrate on the neutrino sector.\\
The  choice  of  the
VEV's of the neutral component of the Higgs fields are as follows:
$$<\Sigma>=\pmatrix{\Sigma_1^0&\Sigma_1^+\cr
                    \Sigma_2^-&\Sigma_2^0}
          =\pmatrix{u&0\cr
                    0&u},
<\phi_\alpha^0>=\pmatrix{0\cr
                         v_\alpha},
<\eta^0>=x.\eqno(5)$$

The VEV's of the bi-doublet and doublet Higgs fields have been fixed
[9, 12] to yield the same
strength of the leptonic and semi-leptonic intractions with $u$ = 3.5 TeV
and $\sqrt{\sum_{\alpha} v_\alpha^2}$ = 125$\sqrt 2$ GeV.
The Higgs potential of the model is discussed in Ref.10 and on
minimization of the potential, the relationship between $x$ and $u$
turns out to be\\
$$u=\gamma x\eqno(6)$$
\noindent
where $\gamma$ is some combination of the coefficient of the
Higgs potential. For $\gamma<< 1$, we obtain the hierarchy of
the VEV's of the Higgs fields as \\
$$x>>u>>v_\alpha.\eqno(7)$$

In a previous paper [10], we have discussed the consequences
of Type B transformation under $S_3\times Z_4$ discrete symmetry
on the neutrino mass.
With
$\gamma\sim {10}^{-3}$, the Yukawa couplings $g_1^\prime$(in the mass terms
($\nu_{eR}\nu_{eR} + \nu_{\mu R}\nu_{\mu R})\eta$)$\sim$ 1
and $f_1^\prime$ ( in the mass terms of $({\bar{l_{1L}}}\nu_{\mu R} +
{\bar{l_{2L}}}\nu_{e R})\tilde\phi_1$) $\sim {10}^{-3}$,
we obtain $m_{\nu_e} = m_{\nu_\mu}
= m_0^\prime$ = 2.8 eV implying $\Delta_{21}$ = 0 and
$\Delta_{32} = [{({\xi_0^2\over \xi_0^\prime})}^2 - 1] {m_0^\prime}^2 =
4\times {10}^{-3} {eV}^2$ ( to explain the atmospheric neutrino anomaly)
where $\xi_0$ = ${g^\prime_2\over g^\prime_1}$, $\xi_0^\prime$
= ${f_2^\prime\over f_1^\prime}$.
$f_2^\prime$ and $g_2^\prime$ are the coefficients of the mass terms
in $({\bar{l_{3L}}}\nu_{\tau R}\tilde\phi_2)$ and
$(\nu_{\tau R}\nu_{\tau R}\eta)$
respectively. Interestingly, the ratio $({\xi_0^2\over \xi_0^\prime})$
determines the departure of the mass of $\nu_\tau$ from $m_0^\prime$.
Thus the LSND data
($\Delta_{21} \sim (0.5- 10) eV^2$) cannot be
explained in the model with Type B discrete symmetry.
However, Type A discrete symmetry can accommodate the LSND
data and we discuss its consequences now.\\

For Type A  discrete symmetry the matrices $m_D$ and $m_R$
are of the form \\

$$m_D=\pmatrix{0&0&f_1 v_1\cr
               0&f_2 v_1&0\cr
               f_1 v_1&0&0}
     =\pmatrix{0&0&a\cr
               0&\xi a&0\cr
               a&0&0}\eqno(8a).$$
where $a= f_1 v_1$ and $\xi = {f_2\over f_1}.$

$$m_R=\pmatrix{g_1 x&0&0\cr
                0&g_2 x&0\cr
                0&0&g_1 x}
      =\pmatrix{b&0&0\cr
                0&\xi^\prime b&0\cr
                0&0&b}\eqno(8b)$$

where $g_1 x = b$ and $\xi^\prime = {g_2\over g_1}$.\\
\noindent
All the right-handed Majorana neutrinos
get masses above the ununification symmetry breaking scale due to the choice
of VEV's given in Eqn.(6).
It may be noted
that the mass matrix $m_R$ is flavour diagonal  and,  hence,  no
transition magnetic moment can arise  at  the  one  loop
level due to ordinary leptons. However, the spectator fermions can
contribute to such magnetic moment of the Majorana neutrinos [10].
Furthermore, the spectator fermions allow non-zero
$\nu_e-\nu_\mu$, $\nu_\mu-\nu_\tau$
mixing angles although these are zero at the tree level.\\

Diagonalization of the  mass  matrix  $M_\nu$  given in  Eqn.(2)  leads to
the following
eigenvalues
$$m_{\nu i}= -{{m_i}^2\over M_i}\eqno(9)$$

where $m_i$'s (i=1,2,3) are the eigenvalues of $m_D{m_D}^T$ and
$M_i$'s are the eigenvalues of $m_{R}$.
Thus, we obtain
the
neutrino mass terms which are given by
$$m_{\nu_1}=m_{\nu_3}= -{a^2\over b} = m_0\eqno(10a)$$
$$m_{\nu_2}= ({\xi^2\over \xi^\prime})m_0\eqno(10b)$$
\noindent
where $m_0 =-{a^2\over b}$.\\

With the previous choice of model parameters
$v_1$=100  GeV, $u\sim$ 3.5TeV, $\gamma\sim 10^{-3}$,
 $g_1\sim 1$,
$f_1\sim {10}^{-3}$,
$m_0$ comes out to be 2.8 eV
as before.
However, the model contains a tiny parameter space and
there is not much freedom in the variation of the model
parameters. In particular, $v_1$ is restricted in the range
$(100- 125\sqrt 2)$GeV and $u >3.5$ TeV in order to be consistent
with the low energy charged current data.\\

The $({\xi^2\over \xi^\prime})$
term lifts  the  neutrino mass degeneracy
that can be determined from the recent
LSND experiment. Thus, we obtain
$({\xi^2\over \xi^\prime})$
in the range
$(0.7- 3.16)$ and
$m_{\nu_2}$ in the range (0.7- 3.16)$m_0$ eV. However, the atmospheric
neutrino anomaly ($\Delta_{23}\sim {10}^{-2}$) can not be explained in the
model with Type A discrete symmetry as by fitting the LSND result
we require $\Delta_{23} = \Delta_{21} \sim (0.5- 10) eV^2$.\\

In summary, we conclude that an extended Harvard Model including
$S_3\times Z_4$ discrete symmetry with three light neutrinos
having mass degeneracy of the order of 2.8 eV between two cannot
accommodate at the same time the
solar neutrino, atmospheric neutrino and the LSND data.\\

\vskip 1cm

A.R wishes to acknowledge the hospitality offered by the ICTP, Trieste,
Italy during his visit where the work was initiated and thanks
Biswajoy Bramhachari for many helpful discussions.
A.G.acknowledges financial support from  the UGC, Govt. of India.
\newpage

\begin{center}
{\Large\bf References}
\end{center}
\begin{enumerate}

\item  K. Lande et. al. Proc. XXVth Int. Conf. on High Energy Physics,
Singapore, ed. K. K. Phua  and  Y.  Yamaguchi,  World  scientific
(Singapore, 1991); K.S.  Hirata  et.al.  Phys.Rev.  Lett.  66, (1991) 9;
P. Anselmann et. al. Phys. Lett. B285, (1992) 376;  A.I.
Abazov et.al. Phys. Rev. Lett. 67, (1991) 3332;  V.Garvin  ,  talk
at Int. Conf. on HEP , Dallas 1992.

\item  E. M. Beier et.al.  Phys. Lett.  B280, (1992) 149;  D. Casper
 et.al. Phys. Rev. Lett. 66, (1991) 2561.

\item  C. Athanassopoulos, et. al. Los Alamos National
Laboratory, Report No. LA- UR- 95- 1238, (April 1995).

\item  A. N. Taylor and M. Rowan Robinson, Nature  359, (1992) 396;
M. Davis , F. J. Summers and D. Schlegel, Nature 359, (1992) 393.

\item  L. Wolfenstein, CMU-HEP 95-05, DOE-ER/ 40682, May 1995.

\item  A. Zee , Phys.Lett. B93 , (1980) 389.

\item  E. Ma and P. Roy, UCRHEP- T145, TIFR/TH/95-17, April, 1995.

\item  H. Georgi,  E. E. Jenkins  and  E. H. Simmons,  Phys.  Rev.  Lett.
  62, (1989) 2789,  63 , (1989) 1504 (E), Nucl. Phys.
  B331, (1990) 541.

\item  A. Ghosal, A. K. Ray and K. Bandyopadhyay  Phys. Rev.
D51, (1995) 1314.

\item  A.Ghosal and  A.K. Ray  (Submitted to Phys. Rev. D).

\item M. Gell Mann, P. Ramond  and R. Slansky in "Supergravity", ed.
by  D. Z. Freedman, (North Holland, 1979), T. Yanagida, in
Proceedings of "Workshop on Unified theory and Baryon Number in
the Universe"  ed. by  O. Sawada and A. Sugamoto,  (KEK, 1979),
R.N. Mohapatra and G. Senjanovic, Phys.  Rev.  Lett. 44, (1980) 912;
Phys. Rev. D23, (1981) 165.

\item  E. Ma and S. Rajpoot, Mod. Phys. Lett. A5, 979, (1990) 1529.
\end{enumerate}
\end{document}